\documentclass[12pt]{iopart}

\usepackage{iopams}
\usepackage{graphicx}
\newcommand{\ket}[1]{\mbox{$\mid \! #1 \, \rangle$}}
\newcommand{\bra}[1]{\mbox{$\langle \, #1 \! \mid$}}

\newcommand{\abs}[1]{\mbox{$\mid #1 \mid$}}

\newcommand{\dg}[1]{#1 \,^{\circ}}
\newcommand{\eea}{\end{eqnarray}}
\newcommand{\bea}{\begin{eqnarray}}
\newcommand{\eeas}{\end{eqnarray*}}
\newcommand{\beas}{\begin{eqnarray*}}

\newcommand{\sx}{\ensuremath{\sigma_x}}
\newcommand{\sy}{\ensuremath{\sigma_y}}
\newcommand{\sz}{\ensuremath{\sigma_z}}

\newcommand{\openone}{\leavevmode\hbox{\small1\normalsize\kern-.33em1}}

\newcommand{\figref}[1]{figure~\ref{#1}}
\newcommand{\tabref}[1]{table~\ref{#1}}

\begin{document}

\title[Quantum teleportation and entanglement swapping]{Quantum teleportation and entanglement swapping with linear optics logic gates}

\author{Christian Schmid$^{1,2}$, Nikolai Kiesel$^{1,2}$, Ulrich K. Weber$^{3}$, Rupert Ursin$^{4}$, Anton Zeilinger$^{4}$ and Harald Weinfurter$^{1,2}$}

\address{$^{1}$Department f\"{u}r Physik,
Ludwig-Maximilians-Universit\"{a}t, D-80797 M\"{u}nchen,
Germany\\
$^{2}$Max-Planck-Institut f\"{u}r Quantenoptik, D-85748 Garching, Germany\\
$^{3}$Materials Department, University of Oxford, Oxford OX1 3PH,
United Kingdom\\ $^{4}$Fakult\"{a}t f\"{u}r Physik, Universit\"{a}t
Wien, A-1090 Wien, Austria} \ead{christian.schmid@mpq.mpg.de}
\begin{abstract}
We report on the usage of a linear optics phase gate for
distinguishing all four Bell states simultaneously in a quantum
teleportation and entanglement swapping protocol. This is
demonstrated by full state tomography of the one and two qubit
output states of the two protocols, yielding average state fidelities of about $0.83$ and $0.77$, respectively. In addition, the
performance of the teleportation channel is characterised by
quantum process tomography. The non-classical properties of the entanglement swapping output states are further confirmed by the violation of a CHSH-type Bell inequality of $2.14$ on average.
\end{abstract}


\maketitle

\section{Introduction}
Quantum teleportation \cite{bennett93} and entanglement swapping
\cite{zukowski93} are fundamental elements of quantum
communication protocols and thus play an important role in a
number of applications. Both processes rely on the projection of
two qubits onto maximally entangled Bell states. As every qubit
can be analysed only separately, this detection requires to map
the four Bell states one-to-one onto four distinguishable,
separable states. Such a
disentangling operation 
can be realised by elementary two-qubit quantum gates, e.g., a
controlled not (\textsc{cnot}) or controlled phase gate
(\textsc{cphase}). Teleportation was demonstrated already with a
number of different systems, where the gate operations can be
achieved (e.g.~\cite{nielsen98, furusawa98, riebe04, barrett04}).
However, while photons doubtlessly are the most proper quantum
system for communication tasks, the implementation of two photon
quantum gates is not straight forward as there is no photon-photon
interaction with reasonable coupling strength\footnote{Quantum
teleportation of a photon polarization state with complete Bell
state analysis was once demonstrated using non-linear
effects, though with vanishingly small probability \cite{shih01}.
Deterministic schemes relying on entanglement in additional
degrees of freedom are not suited for teleportation
\cite{Schuck}.}. Beginning with the initial experiments
\cite{bouwtele, swap}, two photon interference \cite{weinfurter94,
braunstein95} was employed to identify up to two of the four Bell
states and, recently, a probabilistic identification of three Bell
states using POVM-operators was demonstrated \cite{gisin05}. As
introduced by Knill, Laflamme and Milburn (KLM) \cite{KLM01}, all
optical two qubit quantum logic can be achieved near deterministically using linear
optics plus conditioned detection and ancillary qubits. The latter can be omitted when probabilistic gate operation is sufficient \cite{Ralph_PG, HT_PG}. Several
experiments already proved the feasibility of these approaches (see \cite{pan01, pittman02, San04} or \cite{obrien03}, respectively). Recently a significant improvement with respect to reliability and stability of a linear
optics logic gate was reported \cite{pgsimpleII, pgsimple,
pgsimpleIII} which allows to employ such gates in multi-photon
quantum communication protocols. Here we report on the
implementation of quantum teleportation and entanglement swapping
including probabilistic, complete Bell state analysis (BSA)
accomplished by the use of a linear optics \textsc{cphase} gate.

\section{Experiment}
\subsection{The Bell State measurement}\label{sec:bsa}
Let us start with shortly sketching the functionality of the gate \cite{pgsimple}:\\
The operation is defined by 
\begin{equation}
\label{eq:PG} \textsc{cphase}= \left\{
\begin{array}{ccr}
\ket{HH}& \rightarrow & \ket{HH}\phantom{,} \\ 
\ket{HV} & \rightarrow & \ket{HV}\phantom{,} \\
\ket{VH}& \rightarrow & \ket{VH}\phantom{,} \\
\ket{VV}& \rightarrow &-\ket{VV},
\end{array}\right.
\end{equation}
where  the logical 0 and 1 are represented by the linear
horizontal (H) and vertical (V) polarisation states of a photon
respectively. To obtain the $\pi$-phaseshift for the term \ket{VV}
only, the gate-input photons are overlapped on a beam splitter
with polarisation-dependent splitting ratio (PDBS), where the
transmission for horizontal polarisation $T_H=1$, and for vertical
polarisation $T_V=1/3$. As horizontal polarisation is not
affected, no interference can occur for $\ket{HH}$ and the state
does not change. The same holds for $\ket{HV}$ and $\ket{VH}$
where the photons are distinguishable by polarisation and
therefore do not interfere. Only if two vertically polarised
photons are passing the gate two photon interference occurs. For
this case we obtain a $\pi$-phase shift if the ratio of the
amplitudes for both photons being reflected is greater than the
one for both being transmitted. In order to obtain equal
amplitudes for all four output states a transmission $T_V=1/3$ is
required, together with two beam splitters with reversed splitting
ratio ($T_H=1/3$, $T_V=1$) placed after each output of the PDBS
(see \figref{fig:setup}). The gate operation succeeds if one
photon is detected in each of the two outputs of the gate, which
occurs in 1/9 of all cases.

The \textsc{cphase} gate can be used to perform a complete Bell
state projection measurement by mapping four Bell states onto four
orthogonal product states. Considering as input, for example, the
maximally entangled Bell state $\ket{\widetilde{\phi^+}} =
1/\sqrt{2}(\ket{H+}+\ket{V-})$, where $+\,(-)$ denotes $\dg{+45}\,
(\dg{-45})$ linear polarisation, the gate will do the following
operation:
\begin{eqnarray}
\ket{\widetilde{\phi^+}} & = & \frac{1}{2}\;\bigl(\ket{HH}+\ket{HV}+\ket{VH}-\ket{VV}\bigr) \nonumber \\
& \stackrel{{\tiny \textsc{cphase}}}{\rightleftharpoons} &
\frac{1}{2}\;\bigl(\ket{HH}+\ket{HV}+\ket{VH}+\ket{VV}\bigr) \nonumber \\
& = & \frac{1}{2} \; \bigl(\ket{H}+\ket{V}\bigr) \otimes
\bigl(\ket{H}+\ket{V}\bigr)=\ket{++}.
\end{eqnarray}
This means, the gate transforms between the product state
$\ket{++}$ and the maximally entangled Bell state
$\ket{\widetilde{\phi^+}}$, (analogously we obtain for
$\ket{\widetilde{\psi^+}} \rightleftharpoons \ket{+-}$, $
\ket{\widetilde{\phi^-}} \rightleftharpoons \ket{-+}$ and
$\ket{\widetilde{\psi^-}} \rightleftharpoons \ket{--}$).
Consequently, by detecting one of these four product states
\emph{behind} the phase gate, we know that the photons have been
in the corresponding Bell state \emph{before} the phase gate.

For quantum teleportation and entanglement swapping it is, in principle, not necessary to project onto all four Bell states to perform these protocols. The first realizations indeed used the projection onto a single Bell-state only, neglecting the other cases. This results in a success probability of $1/4$ \cite{bouwtele, swap} . The best success probability achieved is $1/2$ and is known to be the theoretical limit when using linear optics without ancillary qubits  \cite{weinfurter94, braunstein95, Lue99, Vai99}. 
Even though we do not neglect any Bell state in our scheme, our success probability is limited by the efficiency of the gate operation, which is $1/9$, and therefore lower than in the other schemes. However, the beauty in the application of the \textsc{cphase} gate is the possibility to detect all four Bell states in a setup just as simple as the (single-state) Bell state projection of the initial demonstration of quantum teleportation \cite{bouwtele}.

Moreover, for quantum teleportation and entanglement swapping one might even mimic the detection of all four Bell states with a setup that detects only a restricted number, by randomly switching between the detected set of states. We like to emphasize, however, that this approach relies only on a statistical mixture of all four Bell states. In contrast, in an \textsc{cphase} gate scheme, a coherent superposition of all Bell states is obtained. This is a fundamental difference and might be crucial for other tasks that rely on the detection of Bell states. Such a situation is present, e.g., in the quantum games scheme, where the referee relies on a quantum gate to analyze entangled states \cite{games}. As discussed before, the referee will, due to the limited success probability of BSA, have to discard several games. But by detecting coherently all four states he does not give the players any chance to cheat, as it might be possible for other BSA implementations.

\subsection{Photon state preparation and detection}
\begin{figure}
\begin{center}
\includegraphics[width=8.5cm,clip]{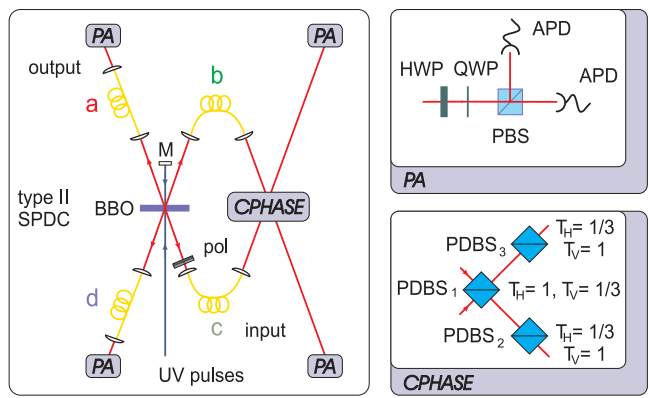}
\caption{\label{fig:setup}\textbf{Experimental setup for the
quantum teleportation and the entanglement swapping experiment,
respectively.} The three and four photon states are provided by
two EPR-pairs originating from type II spontaneous parametric down
conversion (SPDC) processes. UV pulses are used to pump a
$\beta$-Barium Borate (BBO) crystal twice. If a photon pair is
created in each of the two passages, teleportation or entanglement
swapping can be performed. After passing the crystal, the beam is
reflected back by a UV-mirror (M). Half- and quarter wave plates
(HWP, QWP) in conjunction with polarising beam splitters (PBS) are
used for the polarisation analysis (PA). The complete BSA is done
by a controlled phase gate (CPHASE) consisting of three
polarisation dependent beam splitters (PDBS$_1$ - PDBS$_3$). The
photons are spectrally selected with interference filters.}
\end{center}
\end{figure}

In the experiment, the input states for teleportation and entanglement swapping are generated with spontaneous
parametric down conversion. A 2~mm thick BBO ($\beta$-Barium
Borate) crystal is pumped by UV pulses with a central wavelength
of 390~nm and an average power of 700~mW from a frequency-doubled
mode-locked Ti:sapphire laser (pulse length 130~fs). After passing
the crystal the beam is reflected back by a UV-mirror in a
distance of about 3~cm to enable SPDC also into a second pair of
beams. We use degenerate, non-collinear type-II phase matching to
obtain pairs of orthogonally polarised photons at a wavelength of
$\lambda \simeq 780$ nm in the forward and backward direction of
the BBO crystal, respectively. The photons propagating along the
characteristic intersection lines of the emission cones are
coupled into single mode fibers defining the four spatial modes
$a$, $b$, $c$, $d$. The spectral selection is done with narrow
bandwidth interference filters F ($\Delta\lambda = 2$~nm in the
\textsc{cphase} gate and $\Delta\lambda = 3$~nm in modes $a$ and
$d$) before detection. For initial alignment of the spatial
overlap at the partially polarising beam splitter in the
\textsc{cphase} gate we use the two photons of one pair for higher
count rates, whereas the temporal overlap can be aligned via
Hong-Ou-Mandel interference of two independently created, heralded
single photons in the forward and backward direction (see
\figref{fig:setup}). The polarisation states of the photonic
qubits are analysed by half- and quarter wave-plates in
combination with a polarising beam splitter cube and detected by avalanche photon diodes (APD). The setup is
stable over several days with typical detection rates of 180
fourfold coincidence counts per hour. The coincidence count rates
have to be corrected for different detector efficiencies in the
polarisation analysis of modes $a$, $b$, $c$ and $d$, which are
determined relative to each other. The errors on all quantities
are deduced according to Poissonian counting statistics of the
raw detection-events and the detection-efficiencies.

\subsection{Teleportation}

The goal of quantum teleportation is to transfer the most general
polarisation state $\ket{\chi}_c=(\alpha \ket{H}_c+\beta
\ket{V}_c)$ with arbitrary amplitudes $\alpha$, $\beta$ of the
photon in mode $c$ onto the photon in mode $a$. In order to do so,
we need, firstly, a maximally entangled Bell state in modes $a$
and $b$, and, secondly, a complete Bell state projection
measurement between the photons in mode $b$ and $c$. We obtain the
Bell state by proper alignment of the photon pair originating from
the forward downconversion. The photon which will carry the state
$\ket{\chi}_c$ is provided by the backward emission of the down
conversion which is operated 
as a heralded single photon source
with the photon in mode $d$ initialising the trigger. The
polarisation state $\ket{\chi}_c$ can be prepared by a polariser
in front of the fiber coupler in mode $c$ and proper alignment of
the fibre's polarisation controller.

To demonstrate teleportation, we have prepared the states $\ket{H}$,
$\ket{V}$, $\ket{+}$, $\ket{R}=1/\sqrt{2}\;(\ket{H}+\rmi\ket{V})$ as input states
and carried out a single qubit tomography \cite{qutomo} in the output mode
$a$. From this we obtain the density matrix $\rho_{\mathrm{exp}}$
of the experimentally teleported states and can calculate the
fidelities to the input states, $\mathcal{F}_H=0.93 \pm 0.02,\;
\mathcal{F}_V=0.75 \pm 0.05,\; \mathcal{F}_+=0.79 \pm 0.02,\;
\mathcal{F}_R=0.84 \pm 0.03$, with
\begin{equation}
\mathcal{F}_k=\phantom{\langle}_c\bra{\chi_k} \;\mathcal{U}_i
\rho_{\mathrm{exp}} \mathcal{U}^{\dagger}_i \; \ket{\chi_k}_c
\;\;,
\end{equation}
where $k=H,V,+$, and $R$. 

Depending on the outcome of the Bell projection measurement one
has to apply one out of four unitary operations $($represented by
the identity or one of the three Pauli matrices, respectively,
$\mathcal{U}_i=\openone ,\; \sx ,\; \sz ,\; \mathrm{or}\; \rmi
\sy)$ in order to recover the original state in the teleported
mode. Therefore, the fidelities are calculated after application
of the unitary operation on the data and averaging over the four
different results of the BSA.

As can be seen, the quality of the output states differs
            for the various input states. This can be understood
            by considering the influence of imperfect gate
            operation. For the experimental gate the main reason for deviation from ideal performance is caused by
            lack of interference at PDBS$_1$. From the
            considerations in section~\ref{sec:bsa} it can be
            easily seen that for perfectly distinguishable photons
            the probability to obtain a coincidence detection is
            enhanced by a factor of five for the input state $\ket{VV}$. This is
            because if the photons do not interfere, the
            probabilities rather than the amplitudes for both being reflected or both being
            transmitted add up to $(1/3)^2+(-2/3)^2=5/9$.

            Taking
            that into account, it is obvious that the teleportation works best for the state
            $\ket{H}$, as in this instance no
            interference is required. Consequently, from this point of view, the
            output state for the input $\ket{V}$ is expected to be
            the worst. The states $\ket{+}$ and $\ket{R}$ should be
            teleported approximately at the same quality on
            average. However, for the state $\ket{+}$ the fidelity
            of the output state depends on the result of the
            measurement in the \textsc{cphase} gate. The measured fidelities exhibit roughly the expected
            behavior: The loss in quality for $\ket{H}$ is not caused by lack
            of interference but determined
            by impurity of the input states. For $\ket{R}$,
            $\ket{+}$ and $\ket{V}$
            both effects are relevant. However, $\ket{V}$ is
            maximally impaired by the imperfect interference and
            exhibits indeed the lowest fidelity.
            Still, despite all imperfections it is important to note that the
            average fidelities are all well above the optimal
            classical limit of $2/3$.

\begin{figure}
\centering
\includegraphics[width=8.5cm,clip]{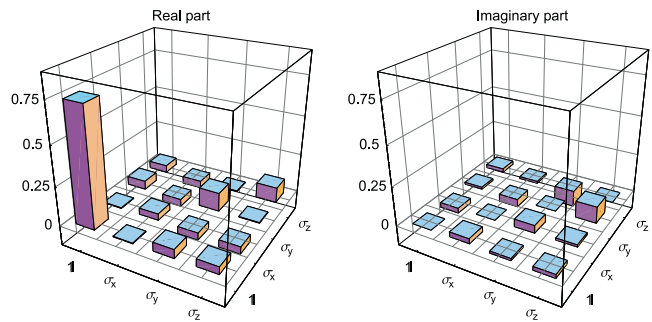}
\caption{\label{fig:teletomo} \textbf{Experimentally reconstructed
teleportation process tomography matrix
$\mathcal{M}_{\mathrm{exp}}$.} As the ideal teleportation process
equals the $\openone$-operation, the height of the $(\openone ,
\openone)$-entry of $\mathcal{M}_{\mathrm{exp}}$ serves directly
as a measure for the fidelity of the experimentally achieved
process. The smallness of the imaginary parts, which are all below
0.1, also confirms the quality of our teleportation procedure.}
\end{figure}

The four chosen input states represent a tomographic set out of
which we can evaluate a teleportation process tomography
\cite{tomo}. From this tomography one obtains the matrix
$\mathcal{M}_{\mathrm{exp}}$ representing the performance of the
teleportation process (see \figref{fig:teletomo}). In this
representation an ideal teleportation
($\mathcal{M}_{\mathrm{theo}}$) corresponds to the identity
operation. Thus the height of the $(\openone , \openone)$-entry of
$\mathcal{M}_{\mathrm{exp}}$ directly gives the so called process
fidelity \cite{profid}:
\begin{equation}
F_p=\mathrm{Tr}[\mathcal{M}_{\mathrm{theo}}\mathcal{M}_{\mathrm{exp}}]
\end{equation}
which is the overlap between the experimentally obtained and the
theoretically expected matrix, and which is the measure for the
quality of the implemented teleportation process. In our
experiment we reached $F_p=0.75$. The limiting factor for the
process fidelity is the
            fidelity of the state which is teleported worst.
            Following the previous discussions
            this is the state $\ket{V}$ for which the output state
            fidelity reaches an average value comparable to
            $F_p$.

\subsection{Entanglement swapping}
The inherent quantum features of the teleportation process are
best seen by performing entanglement swapping. In this quantum
communication method two photons, which have never interacted in
the past, become entangled by teleporting the state of one photon
of an entangled pair onto one photon of another entangled pair. In
the experiment described before, the teleportation of a polarised
photon does not succeed always, e.g. due to experimental
restrictions like limited detection efficiencies. Hence, it could
be argued that the observed teleportation fidelities are a result
of statistical averaging over many measurements. Such arguments
can be directly refuted for entanglement swapping. Here, the
teleported photon is part of an entangled pair, in that sense it
is not polarised. Therefore, the outcome of a measurement on this
photon only is completely random. If the
observed teleportation results for individual one-photon output
states were attributed to statistical averaging, the analogue
experimental procedure would thus unavoidably lead to a random result
for the correlation measurements on two-photon output states. In
the following, however, it will be proven that indeed quantum
correlations can be observed. This confirms the entanglement
contained in the swapped photon pair and proves that teleportation
succeeds for every single instance.

To perform entanglement swapping we start with two entangled
photon pairs, each in the state
$\ket{\phi^+}=1/\sqrt{2}(\ket{HH}+\ket{VV})$ emitted by our down
conversion source in the forward and in the backward direction,
respectively. As before, we accomplish the Bell projection
measurement between modes $b$ and $c$ by the use of the
\textsc{cphase} gate. Consequently, by projecting photons from
these two modes onto a Bell state, the photons from mode $a$ and
$d$
will be left in a maximally entangled state. Which Bell state they
form again depends on the result of the Bell state measurement in
modes $b$ and $c$:
\begin{eqnarray}
\ket{\Psi}_{abcd} =\ket{\phi^+}_{ab} \ket{\phi^+}_{cd}& =
&\frac{1}{2}\,\Bigl(\ket{\widetilde{\phi^+}}_{ad}\ket{\widetilde{\phi^+}}_{bc}
+\ket{\widetilde{\psi^+}}_{ad}\ket{\widetilde{\psi^+}}_{bc} \nonumber\\
&+&\ket{\widetilde{\phi^-}}_{ad}\ket{\widetilde{\phi^-}}_{bc}
+\ket{\widetilde{\psi^-}}_{ad}\ket{\widetilde{\psi^-}}_{bc}\Bigr)
\end{eqnarray}
\begin{figure}
\centering
\includegraphics[width=8.5cm,clip]{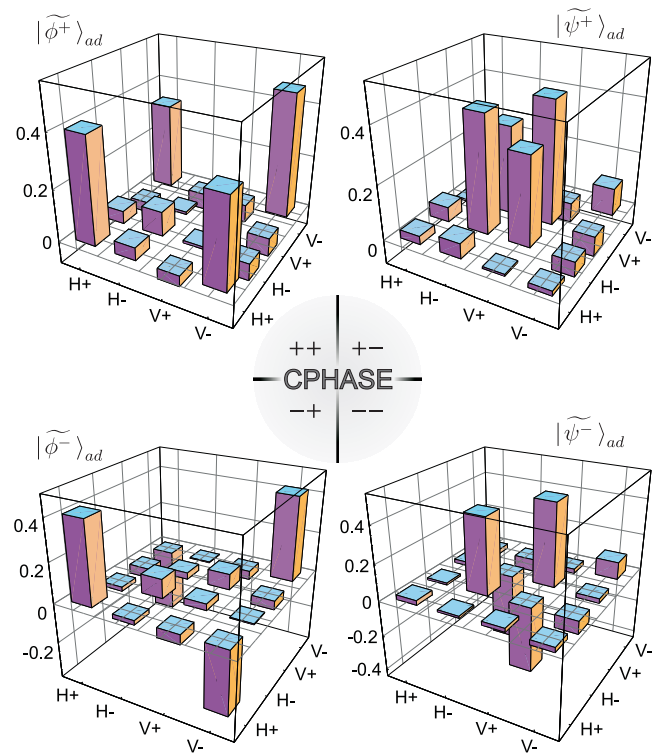}
\caption{\label{fig:swap} \textbf{Experimentally reconstructed
density matrix $\rho_{\mathrm{exp}}$ of the swapped states.}
Different outcomes of the projection measurement in the
\textsc{cphase} gate result in different swapped states. In all
four cases, the four columns, which are significantly different
from noise clearly signal the respective swapped, entangled state
with high fidelity.}
\end{figure}

In order to determine how close the experimentally obtained states
are to the expected ones and whether they are indeed entangled we
performed a two-qubit tomography for photons detected in modes $a$
and $d$ depending on the result in the BSA. From this we obtain
the experimental density matrices $\rho_{\mathrm{exp}}$ (see
\figref{fig:swap}),
out of which we are able to calculate the states' fidelity
$\mathcal{F}$, as well as their logarithmic negativity
$\mathcal{N}$ \cite{neg}. The latter is, as an entanglement
measure, zero for separable states and equal to one for maximally
entangled states. As one can see from \tabref{tab:swap} we get an
entangled state for each of the four Bell state projections in the
\textsc{cphase} gate with fidelities of up to $0.803$ relative to
the corresponding expected Bell state and with an average of 0.773
for all simultaneously detected Bell states.
\begin{table}[t]
\centering
\begin{tabular}{ccc}
Bell state observed & \hspace{0.2cm}Fidelity $\mathcal{F}_{\mathrm{exp}}$\hspace{0.2cm} & Negativity $\mathcal{N}$  \\[1.5ex]

$\ket{\widetilde{\phi^+}}_{cb}$ & \hspace{0.2cm}$0.777 \pm 0.031$\hspace{0.2cm} & $0.660 \pm 0.051$\\
$\ket{\widetilde{\psi^+}}_{cb}$ & \hspace{0.2cm}$0.776 \pm 0.029$\hspace{0.2cm} & $0.666 \pm 0.048$\\
$\ket{\widetilde{\phi^-}}_{cb}$ & \hspace{0.2cm}$0.736 \pm 0.031$\hspace{0.2cm} & $0.582 \pm 0.055$\\
$\ket{\widetilde{\psi^-}}_{cb}$ & \hspace{0.2cm}$0.803 \pm 0.027$\hspace{0.2cm} & $0.720 \pm 0.042$\\[1.0ex]
\end{tabular}
\caption{\label{tab:swap} Experimental results obtained in the
entanglement swapping experiment.}
\end{table}

Quantum teleportation enables efficient communication of quantum
information between remote partners and thus is a core element of
future long distance quantum networks. From that point of view
entanglement swapping is particularly useful, provided one obtains
a swapped state which is entangled strongly enough such as to
exhibit non-local correlations. To check the non-classical
properties of our swapped states we show that they violate a
CHSH-type Bell inequality \cite{chsh}. Using the \textsc{cphase}
gate this can be done at the same time for all four Bell states by
measuring the correlation coefficient:
\begin{equation}\label{eq:chsh1}
\abs{S_\pm}:= \abs{\pm\langle \widehat{A}, \widehat{D} \rangle \mp
\langle \widehat{A}, \widehat{d} \rangle + \langle \widehat{a},
\widehat{D} \rangle + \langle \widehat{a}, \widehat{d} \rangle}
\end{equation}
Herein $\langle \widehat{A}, \widehat{D} \rangle$, $\langle
\widehat{A}, \widehat{d} \rangle$, $\langle \widehat{a},
\widehat{D} \rangle$ and $\langle \widehat{a}, \widehat{d}
\rangle$ are the expectation values of four local operators which
correspond to a polarisation measurement under four sets of
angles; $\dg{0}\,\mathrm{for}\,\widehat{a}$, $\dg{-
22.5}\,\mathrm{for}\,\widehat{D}$, $\dg{-
45}\,\mathrm{for}\,\widehat{A}$ and
$\dg{-67.5}\,\mathrm{for}\,\widehat{d}$, respectively.
$\widehat{A}$, $\widehat{a}$ are acting on qubits in mode $a$ and
$\widehat{D}$, $\widehat{d}$ on qubits in mode $d$.

For local hidden variable models $\abs{S_{\pm}}$ is bounded from above
by $2$.
In our experiment we were able to violate this 
limit for each of the four Bell states
($\ket{\widetilde{\phi^+}}_{ad}: S_{+}=-2.20\pm 0.17$,
$\ket{\widetilde{\psi^+}}_{ad}: S_{+}=2.13\pm 0.15$,
$\ket{\widetilde{\phi^-}}_{ad}: S_{-}=2.12\pm 0.16$,
$\ket{\widetilde{\psi^-}}_{ad}: S_{-}=-2.12\pm 0.18$).
Due to the limited measurement time for each of the four cases the error is relatively high compared to the actual violation.
However, the average value of $2.14 \pm 0.08$ confirms the violation of the Bell inequality.

\section{Discussion and conclusion}
To summarise, we have performed complete BSA in a teleportation
and entanglement swapping experiment by applying a probabilistic,
linear optics \textsc{cphase} gate for photons. The teleported
polarisation states showed fidelities clearly above the classical
bound. The quality of the implemented teleportation and the fact
that we were able to achieve an efficient quantum channel was
confirmed by reconstruction of the quantum process matrix. Running
the entanglement swapping protocol yields high fidelities and
states which are entangled strong enough to violate a Bell
inequality. So, a universal two photon gate based on linear optics was successfully applied for the first time in quantum communication protocols.

Our experiment is a further demonstration
that linear optics gates are no longer feasible just in principle
but have reached a level of functionality and simplicity which
allows their implementation in quantum information applications. The
combination with recently developed active feed-forward techniques
\cite{pre06} will additionally open up new vistas for linear
optics quantum computation.

\ack This work was supported by the DFG-Cluster of Excellence MAP and the European Commission through the EU
Project RamboQ and QAP.\\ R.~U. and A.~Z. acknowledge the support by QCCM.

\end{document}